
\nonstopmode
 \tolerance=10000
\input phyzzx.tex

\def\da {\Delta_}

\REF\zam{A.B. Zamolodchikov, Teor. Mat. Fiz. {\bf 65} (1985) 1205.}

\REF\fatty{V.A. Fateev and S.
Lykanov, \intmp\ {\bf A3} (1988) 507.}
\REF\bais{F. Bais, P. Bouwknegt, M. Surridge and K. Schoutens, \np\ {\bf
B304} (1988) 348, 371.}
\REF\bige {A. Bilal and J.-L. Gervais, \pl\ {\bf
206B} (1988) 412; \np\ {\bf B314} (1989) 646; \np\ {\bf B318} (1989) 579;
Ecole Normale preprint LPTENS 88/34.}
\REF\hee{C.M. Hull,  \np\ {\bf 353
B} (1991) 707.}
\REF\nah{R. Blumenhagen, M. Flohr, A. Kliem, W. Nahm, A. Recknagel and
 R. Varnhagen, Bonn University preprint BONN-HE-90-05(1990).}
\REF\kawat{H. G. Kausch and G. M. T. Watts, \np\ {\bf B354} (1991) 740.}
\REF\pape{H. Lu, C. N. Pope X. Shen and X.J. Wang,
\pl\ {\bf B267}(1991) 356 - 361.}
\REF\nied{M.R. Niedermaier, DESY preprint 91-148.}

\def\np{Nucl. Phys.}
\def\pl{Phys. Lett.}

\def\intmp{Intern. J. Mod. Phys.}

\def\half{{\textstyle {1 \over 2}}}

\Pubnum = {QMW-92-7}
\date = {May 1992}
 \pubtype={}

\titlepage
\title {\bf  SINGULAR CONTRACTIONS OF $W$-ALGEBRAS}

\author {C. M. Hull and L. Palacios\footnote*{supported by CONACyT-MEXICO,
grant No. 52361}}
\
\address {Physics Department,
Queen Mary and Westfield College,
\break
Mile End Road, London E1 4NS, United Kingdom.}


\abstract
{ Many $W$-algebras  (e.g. the $W_N$ algebras) are consistent
for all values of the central charge except for a discrete set of exceptional
values. We show that such algebras can be contracted
to new consistent degenerate
algebras at these exceptional values of the central charge. }

 \endpage
\pagenumber=1

    Many $W$-algebras, such as the
      $W_N$ algebras [\zam - \hee] which are generated by currents
of spin $2, 3,\ldots N $,
have the property that they are defined for all values of the central charge,
except for a discrete set of exceptional values.
It is the purpose of this
paper to investigate such $W$-algebras for these exceptional values of the
central charge.

This work was originally stimulated by
the paper [\pape] which claimed to give a construction of
the $W_4$ algebra for $c=-2$, which appeared to contradict the results of
[\nah, \kawat] who claimed that the $W_4$ algebra was
defined and satisfied the Jacobi identities for all values of $c$ except a
discrete set
of exceptional values, which unfortunately included the value $c=-2$.
This suggested that the $c=-2$ algebra of [\pape] could not be the usual $W_4$
algebra, and raised the question as to what it might be. We will return
to this algebra at the end of the paper.

   The $W_3$ algebra is given by
$$\eqalign{[L_n,L_m] &= {c\over2} {n(n^2 - 1)\over 6}\delta_{n+m}
+ (n-m) L_{n+m} \cr
[L_n,W_m] &= (2n-m)W_{n+m}\cr
[W_n,W_m] &= {c\over3}{(n^3-n)(n^2-4)\over5!} \delta_{n+m}\cr
 &+ {(n-m)\over30}[2(n+m)^2 - 5nm -8]L_{n+m} \cr
 &+ {16\over {22+5c}}(n-m)\Lambda_{n+m}\cr}\eqn\wthree$$
where $$\Lambda_m = \sum_n :L_{m-n}L_n: - {3\over10}(m+3)(m+2)L_m,
\eqn\spfour$$
For any currents $A,B$ with modes $A_n,B_n$,
we define the usual normal ordered product $:A_nB_m:$ by [\zam, \bais].
$$:A_nB_m: = \cases
{A_nB_m, & if $m \ge n$
\cr
B_mA_n, & if $m<n$ \cr}
\eqn\khtkjt$$
The form of the algebra
is uniquely determined by requiring that the Jacobi identity
 be satisfied and that the central charge terms have a conventional
normalisation [\zam]. Note, however, that the structure constants become
divergent if $c = {-22\over 5}$, unless the spin-four quasi-primary field
$\Lambda$ given by \spfour\ vanishes or can be set to zero.
 However, the fact that the form of the coupling constants
 is fixed by requiring a conventional normalisation of the
 currents suggests that it might be possible to absorb the divergence occuring
for $c = {-22\over 5}$ into a rescaling of the generators.
To this end, we consider the redefinition
  $$W_n \rightarrow  W'_n = \sqrt{5c+22} W_n
  \eqn\resca$$
  The new generator $W'_n$ is still a spin-three primary field, but the
  $[W',W']$
  commutator is now
  $$\eqalign{
  [W'_n,W'_m] &= {c  {(22+5c)}\over3}{(n^3-n)(n^2-4)\over5!} \delta_{n+m}\cr
 &+ {(n-m){(22+5c)}\over30}[2(n+m)^2 - 5nm -8]L_{n+m} \cr
 &+ {16}(n-m)\Lambda_{n+m}\cr}\eqn\wwthree$$
For all values of $c$ not equal to $-22/5$, this gives an algebra
which is equivalent to the original $W_3$ algebra. However, for
the rescaled algebra, it is now possible to take the limit
$c \rightarrow {-22\over 5}$ while keeping
$W'$ finite
and obtain a non-singular algebra. The
  resulting algebra has a primary spin-three generator $W'$
  satisfying
  $$
  [W'_n,W'_m] =  {16}(n-m)\Lambda_{n+m} \eqn\wwwthree$$
The absence of any central charge term in this commutator
implies that $W'$ is   null, in the sense
 that the vacuum expectation value
$<W'(z)W'(w)>$ vanishes and $W'_{-n} \mid \phi >$ is a zero-norm
state for any state $\mid \phi >$ satisfying the condition
$W'_{n} \mid \phi >=0$ for $n>0$.
The fact that the algebra generated by $L_n, W'_n$ satisfies   the Jacobi
identities for all $c \ne -22/5$ implies, by continuity, that the
new algebra obtained in the  $c \rightarrow {-22\over 5}$ limit
also satisfies the Jacobi identities.
The construction of this new algebra as a limit of rescaled algebras
is similar to a Wigner-Inonu contraction.

   In general, for a $W$-algebra with generators ${W^A}_n$ of conformal
dimension
$\Delta_A$ (labelled by $A=1,2, \dots$), the algebra will take the form
$$[{W^A}_m,{W^B}_n] = g^{AB} m(m^2-1) \ldots (m^2 -(\Delta_A - 1)^2)
\delta_{m+n} + \dots
\eqn\ghksh$$
(plus $W^A$-dependent terms)
for some constant matrix $g^{AB}$, so that $<W^AW^B> \propto g^{AB}$,
and we shall
refer to   $g^{AB}$ as the metric of the algebra.
On the right hand side of \ghksh, terms involving the
generators have been suppressed. For algebras
with unitary representations $g^{AB}$ should be positive definite and should
vanish if $\Delta_A \neq \Delta_B. $ Then a basis can be chosen to diagonalize
$g^{AB}$ and it is conventional to normalise the generators so that
$$[W^A_m, W^B_n] = {c\over{\Delta(2\Delta-1)!}}\delta^{AB}\delta_{m+n}
m(m^2-1)\ldots (m^2 - {(\Delta - 1)}^2) + \ldots $$
for some central charge $c$.
Given this normalisation, requiring the Jacobi identities to hold fixes the
form
of most of the known   $W $-algebras uniquely. In many cases, the resulting
structure constants are singular for certain discrete values of the central
charge $c$
and in these cases it is again possible
to find a contracted algebra
which is non-singular at these special values of the central charge.
In each case, the contracted algebra has a metric
$g^{AB}$ which has zero eigenvalues and hence includes null generators.

   The $W_4$ algebra is given by [\nah]
$$\eqalign{[{W^A}_n,{W^B}_m]&= {c\over\Delta}
 {m(m^2-1)\ldots (m^2-(\Delta - 1)^2)\over{(2\Delta - 1)!}}
\delta^{AB}\delta_{m+n}\cr
&+ {f^{AB}}_C p_{\da A \da B \da C}(n,m){W^C}_{n+m}\cr
&+ {C^{AB}}_{a} p_{\da A \da B \da a}(n,m){\Lambda^a}_{n+m}\cr}\eqn\wfour$$
where ${W^A}_n ({W^2}_n \equiv L_n, {W^3}_n\equiv W_n, {W^4}_n\equiv V_n)$
are the modes of the
primary fields of spin-$\Delta_A$, ($\Delta_A=2, 3$ and $4$, respectively).
The $p_{\da A \da B \da C}(n,m)$ are certain \lq universal'
 polynomials and the $\Lambda^a$ ($a=1, \dots ,7$) are
 composite
operators  constructed from the primary fields, given by
$$\eqalign{ \Lambda^1= \Lambda &= :W^2 W^2: - {3\over10}\partial^2 W^2 \cr
\Lambda ^2= A &= :W^3 W^2: - {3\over 14}\partial^2 W^3 \cr
 \Lambda^3 =\Gamma &= :W^2 \partial^2 {W^2}: - \partial :W^2\partial W^2:
+ {2\over9}\partial^2 :W^2 W^2:-{1\over42}\partial^4 W^2 \cr
 \Lambda^4= \Delta &= :\Lambda  W^2: - {1\over6}\partial^2 \Lambda \cr
\Lambda^5= \Omega &= :W^4 W^2: - {1\over6}\partial^2 W^4 \cr
\Lambda^6= \Pi &= :W^3 W^3: - {1\over84}\partial^4 W^2
- {40\over9(5c+22)}\partial^2 \Lambda - {5\over36}f^{33}_4\partial^2 W^4 \cr
\Lambda^7= \Theta &= :W^3 \partial W^2: - {2\over5}\partial :W^3 W^2:
+ {1\over 20}\partial^3 W^3 \cr}
\eqn\compog$$
with $:AB:$ indicating the normal ordering of the operators $A$ and $B$
expressed by \khtkjt\ when they are decomposed in terms of their modes
$A_n$ and $B_n$.

 The structure constants
 ${f^{AB}}_C$ and ${C^{AB}}_{c}$
 that depend on the central charge [\nah] are given below:
$$ \eqalign{(f^{33}_4)^2 &= {16(7c+114)(c+2)\over 3(5c+22)(c+7)}    \cr
  (f^{44}_4)^2  &= {27(c^2 + c + 218)^2\over (7c+114)(5c+22)(c+7)(c+2)}\cr
  f^{34}_3 &= {3\over4}f^{33}_4\cr}\eqn\conpri$$
$$\eqalign{C^{33}_\Lambda &= {32\over 5c+22} \cr
C^{44}_\Lambda &= {42\over5c+22}    \cr
C^{44}_\Gamma &= {3(19c - 582)\over 20(7c+114)(c+2)}\cr
C^{44}_\Delta &= {96(9c-2)\over(7c+114)(5c+22)(c+2)} \cr
C^{44}_\Omega &= {36(c^2+c+218)\over (5c+22)(c+2)(c+7)}(f^{44}_4)^{-1}\cr
C^{44}_\Pi &= {45(5c+22)\over 2(7c+114)(c+2)}\cr
C^{34}_A &= {39\over7c+114}f^{33}_4; \qquad  C^{34}_{\Theta}
= {3\over 4(c+2)}f^{33}_4
\cr}\eqn\concom$$
The algebra satisfies the Jacobi identities for all values
of $c$, except for the
isolated points $c= -2, -{22\over5}, -7, -{114\over7}$ [\nah]
(although in [\kawat] it is claimed that the Jacobi identities are satisfied
for all values of the central charge except
$c= -2, -{22\over5}, -7, -{114\over 7}, {1 \over 2},- {68 \over 7},
 - {2 \over 5 }$).
The structure constants become
complex in the intervals $-114/7 \leq c \leq -7 $ and $-22/5 \leq c \leq -2$.

Although the $W_4$ algebra does not exist at these exceptional values of  the
central charge,
we can again define new contracted algebras at these values of $c$.
Consider first rescaling  the spin-4 generator, $V_n$, by
$V_n \rightarrow \sqrt {5}\sqrt{c + 2} V_n$ and then taking the limit
$c \rightarrow -2 $. This gives a contracted algebra
generated by the Virasoro generators and the
spin three and four primary fields $W,V$ satisfying
$$\eqalign{[W_n,W_m] &={-2\over3}{n(n^2-1)(n^2-4)\over 5!} \delta_{n+m}\cr
 &+ {(n-m)\over 30}[2(n+m)^2 - 5nm - 8]L_{n+m}\cr
&+ {4\over3}(n-m)\Lambda_{n+m} + {10\over3}(n-m) V_{n+m}\cr
[V_n,V_m] &={33\over180}(n-m)[(n-m)^2 + nm - 7]V_{n+m}\cr
&+ {(n-m)\over 50}[-{93\over20}\Gamma_{n+m}
- 8\Delta_{n+m} +100 \Omega_{n+m}
 + {27 \over 2}\tilde \Pi_{n+m}]\cr
[W_n,V_m] &= \Theta_{n+m}\cr}\eqn\dos$$
where $\Gamma_n , \Delta_n , \Omega_n , $ and $ \Theta_n $ are given by
\compog\ , and
$\tilde \Pi = :WW: - {1\over84}\partial^4 L -{10\over 27}\partial^2 \Lambda
- {25\over36}\partial^2 V. $

Similarly, it is possible to obtain a contracted $W_4$ algebra
for $c= -114/7$. However, after the contractions has been done, some of the
new
structure constants in the conmutators of the spin three and four primary
fields are not real numbers. After the transformation
$$ W \rightarrow -i \sqrt {14\over 13} W\eqn\newv $$
$$ V \rightarrow 4i \sqrt {42\over5}  V, \eqn\newv$$
the structure constant are all real and
the contracted $W_4$ algebra at $c =- {114\over 7}$ is given by the
following set of commutators (in addition to the Virasoro algebra and the ones
that state the fact that the generators $W$ and $V$ are primary fields)
$$\eqalign {[W_m, W_n] &= p_{334}(m,n)[4V_{m+n} + \half \Lambda_{m+n}] \cr
&- {13\over 7} p_{332}(m,n) L_{m+n}\cr
 &+ {247\over 49} {{m+2}\choose 5} \delta_{m+n}\cr}\eqn\trestres$$
$$[W_m, V_n] = {5\over 4}p_{345}(m,n) A_{m+n}\eqn\trescua$$
$$\eqalign{[V_m, V_n] &= p_{446}(m,n)[{3\over4} \tilde \Pi_{m+n} + {1\over 8}
\Delta_{m+n}
- {39 \over 80}\Gamma_{m+n}]\cr
&- p_{444}(m,n) {50\over 3}\cdot {275\over 7} V_{m+n}\cr}\eqn\cuacua$$
where
 $$\tilde \Pi = :WW: - {5\over9}\partial^2V
- {5\over72} \partial^2 \Lambda + {13 \over 1176} \partial^4 L.$$
	After the rescaling $W \rightarrow (c+7)^{-q} W$ and
 $V \rightarrow({\sqrt3} (c+7))^{-p} V $, to obtain a well defined limit
of the structure constants at $c=-7$,
the exponents $p$ and $q$ must satisfy the following relations
$$\eqalign{p &= q  \qquad \hbox{with} \qquad  q \geq \half \cr \hbox{or} \qquad
p&= 2q - \half ; \qquad q > \half \cr}\eqn\siete$$
The first relation  with $q=p=\half$ gives rise to the following commutation
relations
$$\eqalign { [W_n,W_m] &= {10\over3}(n-m) V_{n+m} \cr
[V_n,V_m] &= {n-m\over3} [(n-m)^2 + nm - 7] V_{n+m} \cr
&+ {n-m\over2}[4 \Omega_{n+m}
+ {9\over10} \tilde \Pi_{n+m}] \cr
[W_n,V_m] &= {15\over84}(5n^3 - m^3 -5mn^2 + 3m^2n -17n + 9m) W_{n+m} \cr
&+{12\over5}(3n+2m) A_{n+m} - 3\Theta_{n+m}\cr}\eqn\conuno$$
where $\tilde \Pi = :WW: - {25\over 27} \partial^2 V.$
 If $p=q > \half $ then the contracted commutation relations are
$$\eqalign{[W_n,W_m] &= 0 = [W_n,V_m] \cr
[V_n,V_m] &= {9\over 20}(n-m) \tilde \Pi_{n+m}\cr}\eqn\condos$$
where $\tilde \Pi = :WW:.$

The second relation in \siete\ allows the following limit
$$ \eqalign {[V_n,V_m] &= 0 = [W_n, V_m]\cr
 [W_n,W_m]&= {10\over3}(n-m)V_{n+m}.\cr}\eqn\contres$$
    The contraction for the case $c=-{22\over 5}$ can be obtained
through the rescaling $W \rightarrow (5c - 22)^{-q/2} W$ and $V \rightarrow
(5c - 22)^{-p/2} V$ with
 $(p,q) \in \{ (p,q) / p \geq \half, q \geq \half, p \geq q - \half \quad
 \hbox{and} \quad
p \leq 2q - \half \}.$ The result is a set of null field commutation relations
for the generators of conformal
dimension three and four, with some of the coefficients complex.

To summarise,
we have seen that $W-$algebras that are consistent for all but a discrete
set of exceptional values of the  central charge can be contracted to algebras
that can be continued consistently to these values of $c$, resulting in new
algebras at these $c$-values.
These new  algebras have a degenerate metric, so that some of the
generators have become null and the rank of the algebra has been reduced.

We now return to the paper [\pape], in which  $W_N$ algebras for $c=-2$
are claimed to have been constructed. For the $N=4$ case, for example,
quasi-primary currents $T,W,V$ of spins $2,3$ and $4$ are constructed from free
fermions and it is claimed that they satisfy a closed algebra
which also satisfies
the Jacobi identities and contains the Virasoro algebra with $c=-2$.
 It is usually assumed in such
cases that each of the
quasi-primary currents of spins greater than $2$
 can be modified by expressions
involving composites constructed from the currents and their
derivatives in such a way as to obtain primary currents. The algebra in
this primary
basis should then again be closed and satisfy the Jacobi identities.
If this were the case, and if all the currents were non-null,
then the algebra should be the standard $W_4$ algebra
given by \wfour\ for some value of the central charge. However,
the Virasoro generators are unmodified and so still
satisfy the Virasoro algebra
with $c=-2$, implying that the whole $W_4$ algebra would have $c=-2$.
However, as we have seen, the standard $W_4$ algebra does not exist at $c=-2$,
so it follows that at least one of our assumptions must be wrong.
The quasi-primary fields in the original algebra of [\pape] were all
non-null, \ie\ the expectation values $<TT>$, $<WW>$ and $<VV>$
were all non-zero.
This suggests that the problem must come
from the change of basis form quasi-primary to primary currents, and that
some of the primary currents become null. We now show that this is indeed
the case.
The spin-three current $W$ of [\pape] is in fact primary and needs
no modification, but the spin-four current $V$ is not primary. It is
straightforward to show that
the
current
$$ V'=V-:TT: - {3 \over 10} \partial ^2 T \eqn\yoyo$$
is in fact primary, so that   a primary basis for the
algebra is given by $T,W, V'$. However, for the fermionic realisation of
[\pape], the
current $V'$ vanishes identically, so that the algebra collapses to the
$W_3$ algebra at $c=-2$, which is of course a special case of the contracted
$W_4$ algebra at $c=-2$
constructed above (and it is  possible formally to re-introduce
a null primary spin-four current $V'$  to obtain precisely this
 contracted algebra).  Similarly, for each of the \lq $W_N$' algebras
 of [\pape] with $N>3$, there is again the problem that
 none of the $W_N$ algebras with $N>3$ exist for $c=-2$. The
resolution is again that, on changing to a basis of primary
currents,  each of the currents of spin
greater than $3$ vanish identically, so that for each case the algebra that
is constructed is just $W_3$
 (or  a singular contraction of $W_N$) and so the problem that is avoided.

After this work was completed, we learned of related work by
Niedermaier [\nied],
in which he considers $W_N$
algebras in a basis of quasi-primary currents, and shows that
these algebras are consistent for all values of $c$, and in particular
the structure constants are non-singular functions of $c$. However, the
rank of the algebra (essentially the number of non-null generators)
which is $N-1$ for generic values of $c$, becomes smaller for
special values of $c$, e.g. for $W_4$, the rank
of the algebra becomes less than three for $c=0,-2,-7,-114/7$.
Further, the change of basis from quasi-primary
to primary currents which is possible for generic values of
$c$, becomes singular at $c=-2,-22/5,-7,-114/7$.
By relaxing the conventional
normalisation condition on the primary fields, it is possible
to obtain  a non-singular change of basis to obtain one
of the contracted algebras described above for
$c=-2,-7,-114/7$ but this is not possible for the value $c=-22/5$.
Thus the $W_4$ algebra  of [\nied] is equivalent to the standard $W_4$
algebra for non-exceptional values of the central
charge but is equivalent to the contracted $W_4$ algebras described above.

{\bf Aknowledgments} We would like to thank Horst Kausch and Chris Pope
for discussions and one of us (LP) would like to thank CONACYT-M\'exico for
financial support.

\refout
\bye
\end